\documentclass{elsart}
\usepackage{amssymb}
\usepackage{epsfig}

\newcommand{\be}{\begin{equation}}
\newcommand{\ee}{\end{equation}}
\newcommand{\bea}{\begin{eqnarray}}
\newcommand{\eea}{\end{eqnarray}}

\begin{document}

\begin{frontmatter}

\title{\bf Ferrimagnetism and compensation points in a decorated 3D Ising models}
\author{J. Oitmaa\cite{byline1} and Weihong Zheng\cite{byline2}}
\address{School of Physics,
The University of New South Wales,
Sydney, NSW 2052, Australia.}



\begin{abstract}
We give a precise numerical solution for decorated Ising models on the
simple cubic lattice which show ferromagnetism, compensation points, and
reentrant behaviour. The models, consisting of $S={1\over 2}$ spins on a simple
cubic
lattice, and decorating $S=1$ or $S=\frac{3}{2}$ spins on the bonds,
can be mapped exactly onto the normal spin-$\frac{1}{2}$
 Ising model, whose properties are well known.
\end{abstract}

\end{frontmatter}


Ferrimagnets are materials where ions on different sublattices have opposing
magnetic moments which do not exactly cancel even at zero
temperature.\cite{wol61} An intriguing possibility
then is the existence of a compensation point, below the Curie temperature,
where the net moment changes sign. This has obvious technological significance.

There has been considerable work in recent years in studying
these phenomena through simple models\cite{gon85,lip95,bue97,kan96,dak98,dak00},
where treatments beyond mean-field theory are possible. Of particular interest
are decorated systems, which can be mapped exactly onto simpler models, and
in this way solved either exactly or to a high degree of numerical precision.

The study of decorated Ising model or, more generally, of
Ising model transformations has a long history\cite{fis59,syo}.
Kaneyoshi\cite{kan96} introduced a mixed spin Ising model on the square
lattice with spins $S_A=\frac{1}{2}$ on the vertices, coupled antiferromagnetically
to spins $S_B>\frac{1}{2}$  decorating the bonds. Within the framework of an
effective-field theory with correlations, a number of interesting results were
obtained. In particular, it was found that for $S_B=1$ and a negative crystal field
anisotropy term the system could show up to three separate phase transitions and one
or two compensation points. The crystal field term and a residual
ferromagnetic coupling between the $A$ spins were crucial for these effects.
On the other hand there was no reentrant behaviour and at most
one compensation point for $S_B=\frac{3}{2}$. Dakhama\cite{dak98}
obtained an exact solution for this same system, and confirmed the
existence of compensation points, as predicted by Kaneyoshi\cite{kan96},
but found no evidence for multiple phase transitions.

\begin{figure}[ht] 
\vspace{5mm}
\centerline{\hbox{\psfig{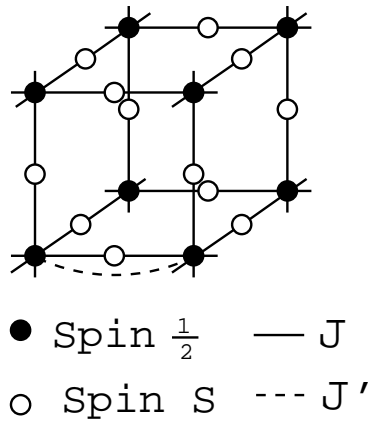}}}
\par
\caption{The mixed spin cubic lattice.
}
\label{fig_1}
\end{figure}

It is quite straightforward to extend this analysis to the simple cubic
lattice, and this is the substance of our paper. Indeed it seems surprising
that this has not been done before, but to our knowledge it has not.
We consider a simple cubic lattice with $N$ spins
$S_A=\frac{1}{2}$ on the vertices and $3N$ spins $S_B>\frac{1}{2}$ decorating
the bonds. The model and its interactions are shown in Figure 1. The Hamiltonian is taken as
\be
H = J\sum_{\langle ij\rangle} \sigma_i S_j - J'\sum_{\langle k l\rangle} \sigma_k \sigma_l
- \Delta\sum_i S_i^2
\ee
where $\sigma_i =\pm\frac{1}{2}$, $S_i = -S, -S+1, \cdots, S$ and $J, J'>0$.

Summing over the B spins we can write the Boltzmann weight for each bond as
\be
\sum_{\mu=-S}^{S} \exp[ K' \sigma_1 \sigma_2 - K \mu (\sigma_1 + \sigma_2) + D \mu^2 ]
= \Lambda \exp (\tilde{K} \sigma_1 \sigma_2)
\ee
where we have introduced coupling constants $K=\beta J$,
$K'=\beta J'$, $D=\beta \Delta$. Thus the model transforms to a spin-$\frac{1}{2}$
Ising ferromagnet on a simple cubic lattice, with coupling constant
\be
\tilde{K} = K' + 2 \ln(\Sigma_1/\Sigma_2)
\ee
with
\be
\Sigma_1 = \sum_{\mu} e^{D\mu^2} \cosh K\mu,\quad
\Sigma_2 = \sum_{\mu} e^{D\mu^2}
\ee

The prefactor $\Lambda$, which is needed to compute the free energy, is given by
$\Lambda^2 = \Sigma_1 \Sigma_2$. Note that the new coupling constant $\tilde{K}$ is
independent of the sign of $J$, and hence our conclusions regarding phase transitions
hold equally well for ferromagnetic nearest neighbour coupling. Of course, in that case,
the sublattice moments will be parallel and there will be no compensation point.

The critical point of the nearest neighbour spin-$\frac{1}{2}$ Ising model on the simple cubic
lattice is known to high accuracy\cite{but} as
\be
\tilde{K}_c = 0.88664
\ee
(note that we use $\sigma=\pm \frac{1}{2}$, rather than $\sigma=\pm 1$)
and hence our model has a critical line, which
can be expressed as
\be
e^{K'/2} \sum_{\mu} e^{D\mu^2} \cosh K\mu = 1.5579 \sum_{\mu} e^{D\mu^2} \label{eqKc}
\ee
For the two cases which we consider explicitly, namely $S=1,\frac{3}{2}$ we can write this is
\bea
&&S=1:\quad e^{K'/2} (1+2e^D \cosh K ) = 1.5579 (1+2 e^D) \label{eq6} \\
&&S=\frac{3}{2}:\quad e^{K'/2} (\cosh {K\over 2} + e^{2 D} \cosh {3 K\over 2} )
=1.5579 (1+e^{2D}) \label{eq62}
\eea
These equations must be solved numerically, and the solutions are discussed below.
We note, of course, that Eqns. (\ref{eqKc})-(\ref{eq62}) have the same form
for any lattice, with the appropriate $e^{\tilde{K}_c/2}$.

The spontaneous magnetizations of the two sublattices are easily obtained. For
the A sublattice we have simply
\be
\langle \sigma \rangle = \frac{1}{2} M_0 (\tilde{K})
\ee
where $M_0(K)$ is the magnetization of the spin-$\frac{1}{2}$ Ising model on the simple cubic
lattice, which can be obtained numerically to high accuracy. The
magnetization of the B sublattice is given by (precisely as in
\cite{dak98})

\be
\langle S \rangle = 2 \gamma <\sigma >
\ee
where
\be
\gamma = \Sigma_3/\Sigma_1
\ee
and
\be
\Sigma_3 = \sum_{\mu} \mu e^{D\mu^2} \sinh K \mu
\ee

The total magnetization is then given by
\bea
M/N &=& \langle \sigma \rangle - 3 \langle S \rangle \nonumber \\
&=& (1-6 \gamma ) \langle \sigma \rangle
\eea
whence any compensation point must lie on the line $6\gamma =1$. Note that this line
depends on $D$ but not on the second neighbour interaction $K'$.
Explicitly for $S=1, \frac{3}{2}$ the compensation point line is
\bea
&S=1:\quad &12 e^D \sinh K = 1 + 2e^D \cosh K  \label{eq14} \\
&S=\frac{3}{2}:\quad
&6 \sinh {K\over 2} + 18 e^{2D} \sinh {3 K\over 2} =\cosh {K\over 2} + e^{2D} \cosh {3K\over 2} \label{eq15}
\eea

We now turn to the results, which we stress are exact to within the very small numerical
uncertainty in $\tilde{K}_c$.

\begin{figure}[ht] 
\vspace{5mm}
\centerline{\hbox{\psfig{figure=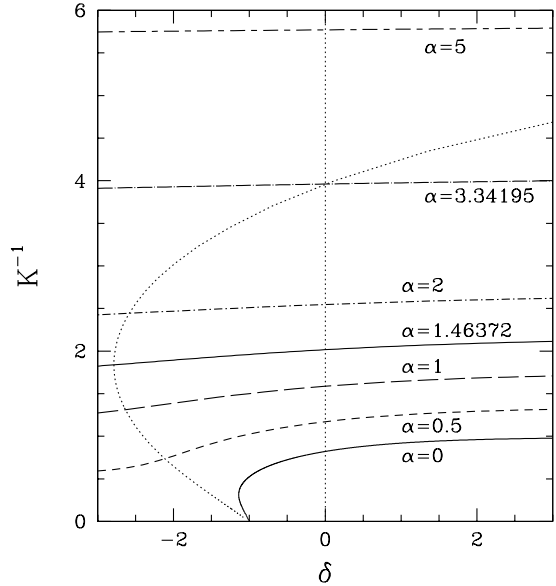,width=8cm}}}
\par
\caption{The phase diagram of the mixed spin cubic lattice
for $S_B=1$. Lines give critical temperature versus single ion anisotropy
for various strengths $\alpha$ of the second-neighbour interaction. The
dotted line is the compensation point line (Eq. \protect\ref{eq14}).
}
\label{fig_2}
\end{figure}

We first consider the case $S=1$. Figure 2 shows critical lines in the
$K^{-1} - \delta$ plane ($\delta = D/K$) for various values of $\alpha = K'/K$.
We see, as expected, that increasing the second-neighbor coupling increases
the critical temperature as does increasing the anisotropy parameter
$\Delta$. The critical line for $\alpha=0$ shows reentrant behaviour, i.e. two phase transitions,
for a small range $-1.14046 <\delta < -1.0$. At very low temperatures the system will be
disordered, since $\Delta<0$ favors that state $S=0$ at B sites. As the
temperature increases the system orders at a lower critical temperature $T_{C_1}$,
passes through an ordered phase, and then disorders again at an upper critical
temperature $T_{C_2}$. Any nonzero second neighbor interaction will give rise ordering at
$T=0$ and at low temperatures, and hence there will be a single critical point (subject
to the caveat below).
Reentrant behaviour of this kind was found in the approximate treatment of the
square lattice\cite{kan96}, but was shown to be spurious
by the exact solution\cite{dak98}. For the simple
cubic lattice it is a real effect.

The compensation point line is shown in Figure 2 as a dotted line.
If this line lies below the phase transition line a
compensation point will be present. Thus for $\alpha=0$ these
is no compensation point, i.e. a second-neighbour
interaction is essential. For $\delta < -2.7858$ no compensation point is possible, while for
$-2.7858<\delta < -1$ two compensation points will occur for
sufficiently large $\alpha$. For $\delta>-1$ a single
compensation point will occur for sufficiently large $\alpha$.

\begin{figure}[ht] 
\vspace{5mm}
\centerline{\hbox{\psfig{figure=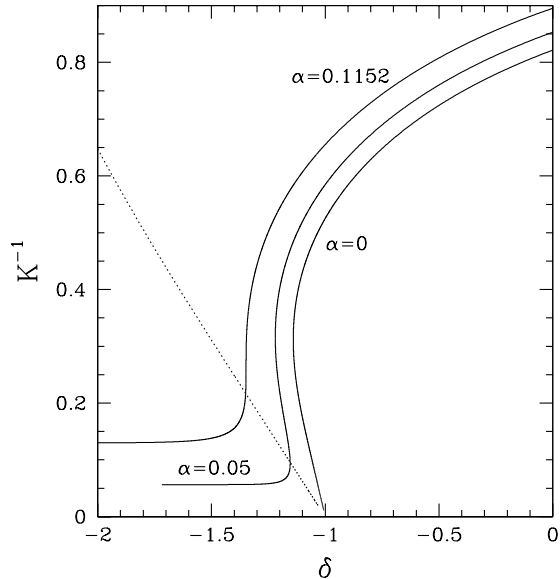,width=8cm}}}
\par
\caption{Critical temperature versus anisotropy for small $\alpha$. In the range
$0<\alpha<0.1152$ the system will have three critical points.
The dotted line is the compensation point line (Eq. \protect\ref{eq14}).
}
\label{fig_3}
\end{figure}

Careful analysis of eqn.(\ref{eq6}) for
small $\alpha$ reveals even more interesting behaviour. In Figure 3 we show critical
temperature curves in the vicinity of $\delta =-1$
for several $\alpha$ values. As is seen, for $0<\alpha < 0.1152$ the curve is
S-shaped, and hence the system has three critical points, going from an ordered
state to a disordered state, then back to an ordered state and eventually to a
high-temperature disordered state.
 While such behaviour has been
seen before in approximate treatments, as for example in the square lattice by
Kaneyoshi\cite{kan96}, we are unware of any cases where it has been
demonstrated in an exact treatment.

\begin{figure}[ht] 
\vspace{5mm}
\centerline{\hbox{\psfig{figure=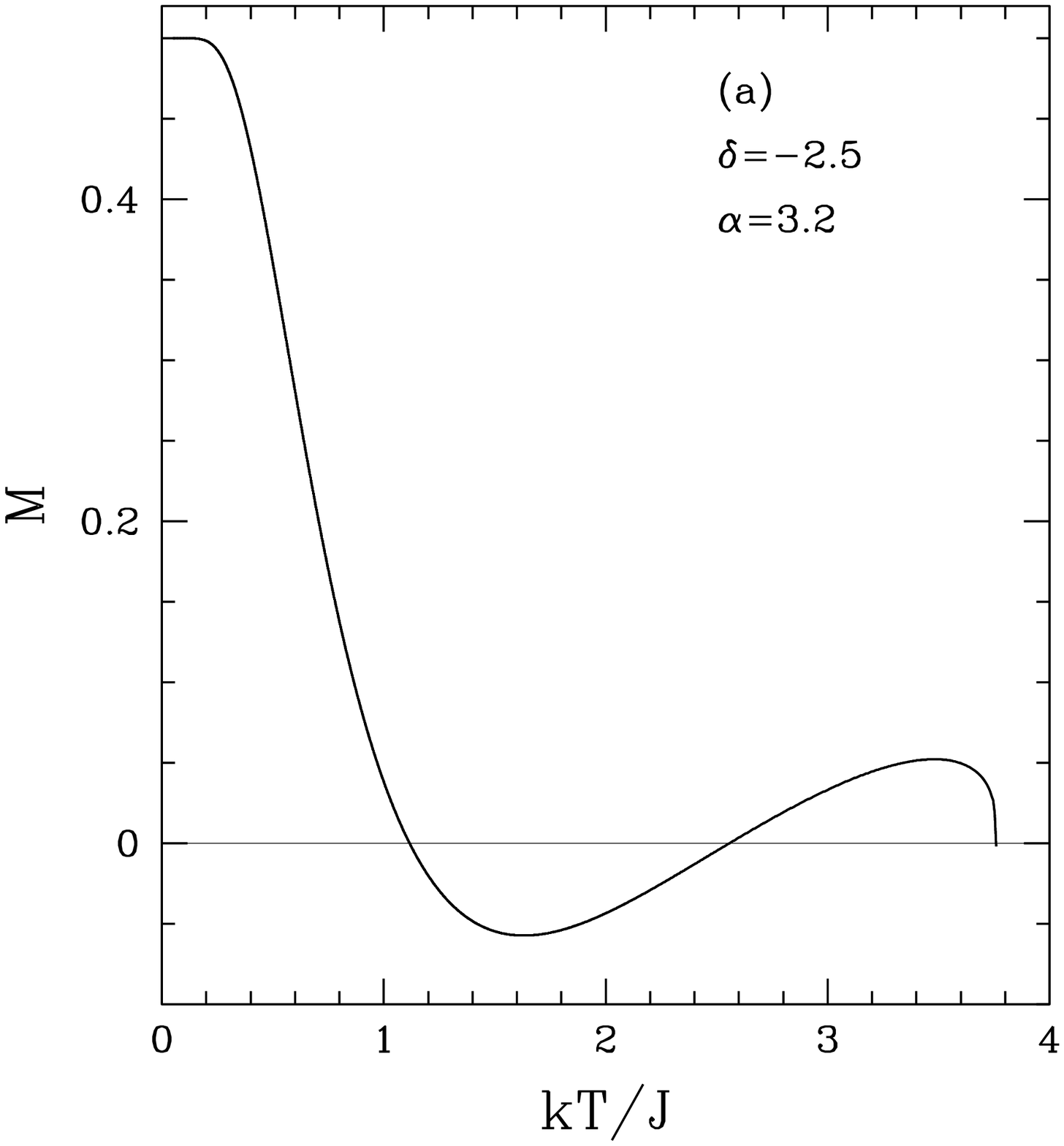,width=7cm}\psfig{figure=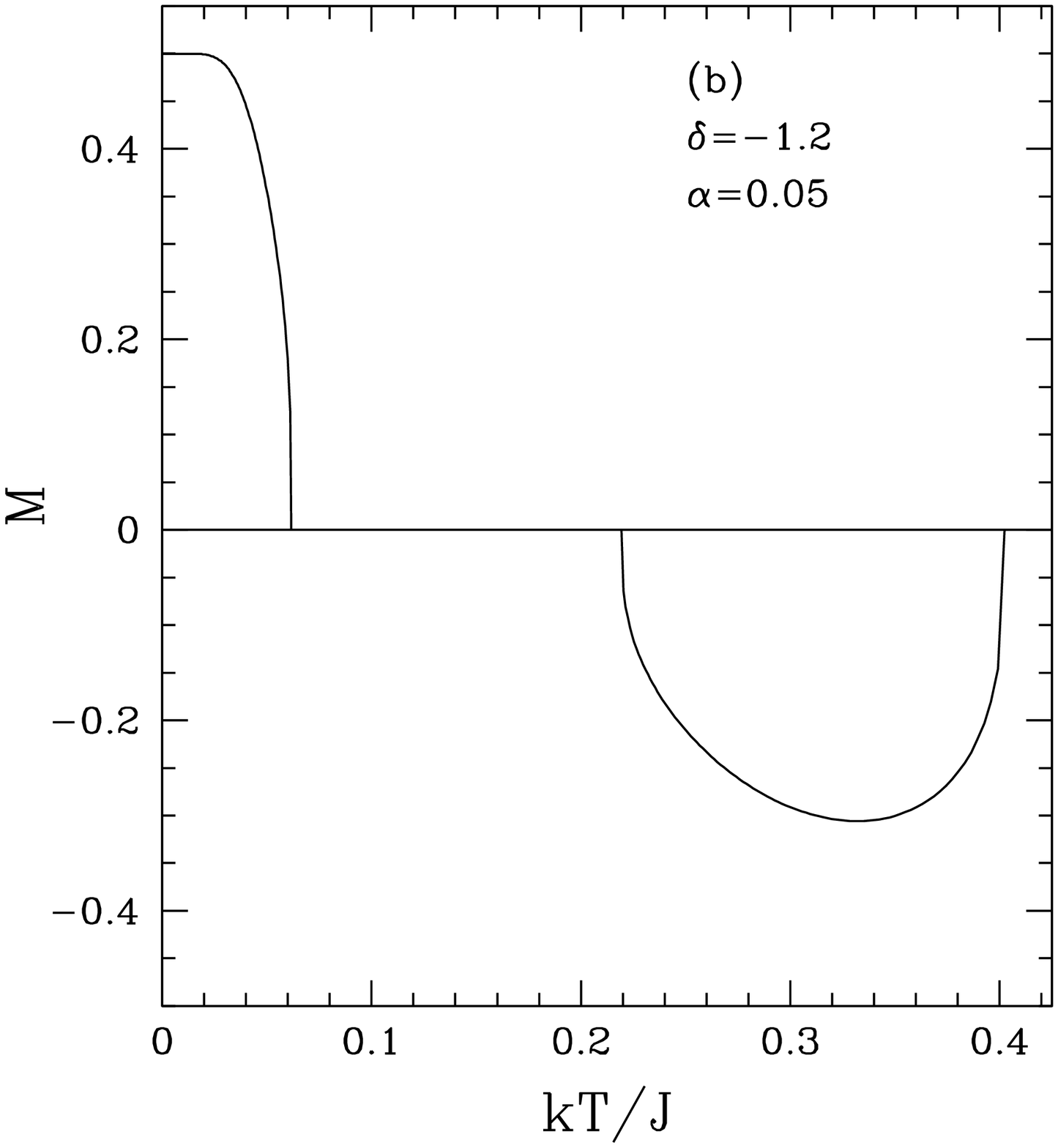,width=7cm}}}
\par
\caption{(a)Total magnetization versus temperature for $\alpha=3.2$,
$\delta=-2.5$, showing two compensation points; (b)Total magnetization versus temperature
for $\alpha=0.05$, $\delta=-1.2$, showing three critical points.
}
\label{fig_4}
\end{figure}

\begin{figure}[hb] 
\vspace{5mm}
\centerline{\hbox{\psfig{figure=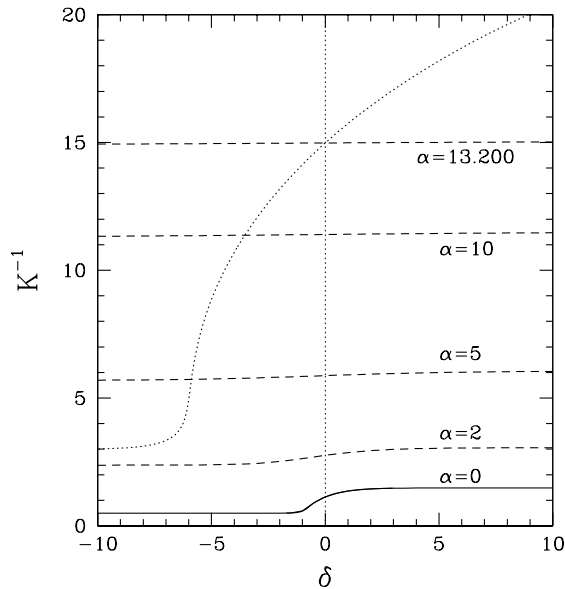,width=8cm}}}
\par
\caption{The phase diagram for $S_B=\frac{3}{2}$. Lines give critical temperature
versus single ion anisotropy for various strengths $\alpha$ of the second-neighbour
interaction. The dotted line is the compensation point line (Eq. \ref{eq15}).
}
\label{fig_5}
\end{figure}

It is useful to compute the total magnetization as a function
of temperature, for particular $(\alpha, \delta)$, to display some of the features
discussed above. The magnetization $M_0 (K)$ for the simple cubic
lattice is not known exactly, but can be computed numerically to high accuracy using the existing
low-temperature series\cite{dom} and the best estimate of the critical behaviour\cite{gui98}.
In Figure 4(a) we shows magnetization versus temperature for the case
$\alpha=3.2$, $\delta=-2.5$. The two compensation points are evident.
Figure 4(b) shows magnetization versus temperature for
$\alpha=0.05$, $\delta=-1.2$. This case has three critical points.
The compensation point line falls in the intermediate disordered phase, and
hence there is no compensation point for this case.
There are parameter values where the compensation point line
falls in the lower $T$ ordered phase, giving three
critical points and a compensation point: an example of this
is for $\alpha=0.11$, $\delta=-1.3343$, where we have three
critical points at $K=3.12055$, 4.18638 and 4.60271, respectively,
while the compensation point is at $K=4.81408$, inside the lower
$T$ ordered phase.

Figure 5 shows the critical point curves and compensation point curve
for $s=\frac{3}{2}$. There is a clear qualitative
difference between the two cases $S=1$ and $S=\frac{3}{2}$, as
previously seen for the square lattice\cite{dak98}. In the
$S=\frac{3}{2}$ case there is no compensation point for
$\alpha<2.5519$, for any $\delta$, whereas a single compensation point
occurs for any $\alpha > 2.5519$,
for a range of $\delta$ values. In all cases there is a single phase transition point.

In conclusion, we have extended the analysis of Ref.\cite{dak98}
to the case of mixed spin Ising models on a decorated simple
cubic lattice. Using the numerical precise value of the critical
point for this lattice we obtain essentially exact results
for phase transitions and compensation points. For the
$S=(\frac{1}{2}, 1)$ system we find reentrant behaviour, with the possibility of
three, two and one phase transition and zero, one, or two
compensation points. For the $S=(\frac{1}{2}, \frac{3}{2})$ system the behaviour
is qualitatively different, with always a single transition and
zero or one compensation point. This difference can be qualitatively understood
through population of the $S=0$ state at low temperatures,
and we expect it to persist for higher integer and half-integer spin values.

For the models studied here, the presence of second-neighbour interaction
is essential for the occurrence of a compensation point. The same conclusion,
for a different mixed spin Ising model, was noted in Ref.\cite{bue97}.

It is interesting to note that our (essentially) exact results are
qualitatively similar to those obtained by a correlated effective
field theory for the square lattice\cite{kan96}. Presumably
the reduced fluctuations in 3-dimensions make the effective-field
theory more reliable than it is for the square lattice.



\end{document}